\documentclass[prb,twocolumn,showpacs,superscriptaddress]{revtex4-1}

\usepackage{amsmath}
\usepackage{amssymb}

\usepackage{longtable}%
\usepackage{dcolumn}
\usepackage{bm}
\usepackage{color}

\usepackage{graphicx}
\usepackage{dcolumn}
\usepackage{bm}
\usepackage{amsmath}
\usepackage{tabu}
\usepackage{multirow}
\usepackage[latin1]{inputenc}
\usepackage{subfigure}
\usepackage{braket}
\usepackage{soul}
\setstcolor{red}
\usepackage{xcolor}
\usepackage{verbatim}
\usepackage[dvipdfm, pdfstartview=FitH, CJKbookmarks=true, bookmarksnumbered=true,
bookmarksopen=true, colorlinks, pdfborder=001, linkcolor=blue, anchorcolor=blue, citecolor=blue]{hyperref}
\newcommand{\sr}[1]{\textcolor{black}{#1}}

\begin{document}
\preprint{Regular article}



\title{Influence of disorder on the signature of $pseudogap$ and multigap superconducting behavior in FeSe}

\author{Sahana R\"o{\ss}ler}
\email{roessler@cpfs.mpg.de}
\affiliation{Max Planck Institute for Chemical Physics of Solids,
N\"othnitzer Stra\ss e 40, 01187 Dresden, Germany}
\author{Chien-Lung Huang}
\altaffiliation{Present address: Department of Physics and Astronomy, Rice University, Houston Texas 77005, US.}
\affiliation{Max Planck Institute for Chemical Physics of Solids,
N\"othnitzer Stra\ss e 40, 01187 Dresden, Germany}
\author{Lin Jiao}
\affiliation{Max Planck Institute for Chemical Physics of Solids,
N\"othnitzer Stra\ss e 40, 01187 Dresden, Germany}
%
%
%
\author{Cevriye~Koz}
\affiliation{Max Planck Institute for Chemical Physics of Solids,
N\"othnitzer Stra\ss e 40, 01187 Dresden, Germany}

%
%
%
%
%
\author{Ulrich Schwarz}
\affiliation{Max Planck Institute for Chemical Physics of Solids,
N\"othnitzer Stra\ss e 40, 01187 Dresden, Germany}
%
%
%
%
\author{Steffen Wirth}
\email{wirth@cpfs.mpg.de}
\affiliation{Max Planck Institute for Chemical Physics of Solids,
N\"othnitzer Stra\ss e 40, 01187 Dresden, Germany}

\date{\today}

\begin{abstract}

We investigated several FeSe single crystals grown by two different methods by utilizing experimental techniques namely, resistivity, magnetoresistance, specific heat, scanning tunneling microscopy, and spectroscopy. The residual resistivity ratio (RRR) shows systematic differences between samples grown by chemical vapor transport and flux vapor transport, indicating variance in the amount of scattering centers.
\sr{Although the superconducting transition temperature $T_c$ is not directly related to RRR}, our study evidences subtle differences in the features of an incipient ordering mode related to a depletion of density of states at the Fermi level. \sr{For instance, the onset temperature of anisotropic spin-fluctuations at $T^* \approx 75$~K, and the temperature of the opening-up of a partial gap in the density of states at  $T^{**} \approx 30$~K are not discernible in the samples with lower RRR.} Further, we show that the functional dependence of the electronic specific heat below 2 K, which allows to determine the nodal features as well as the small superconducting gap, differs significantly in crystals grown by these two different methods.
Our investigation suggests that some of the controversies about the driving mechanism for the superconducting gap or its structure and symmetry is related to minute differences in the crystals arising due to the growth techniques used and the total amount of scattering centers present in the sample.

\end{abstract}


\maketitle
\section{Introduction}

The binary compound FeSe\cite{Hsu2008} belonging to the family of Fe-based superconductors is a fascinating material. Unlike the Fe-pnictide superconductors, FeSe displays an orbital-dependent electron correlation \cite{Aich2010}. As a result, the experimentally obtained band structure close to the Fermi level from the angle resolved photoemission spectroscopy (ARPES) \cite{Mal2014,Wat2015,Evt2016} and quantum oscillation \cite{Wat2015,Tera2014} experiments strongly deviates from that calculated using the density functional theory (DFT)\cite{Sub2008}. At temperature  $T_s\approx$ 87~K, FeSe undergoes a symmetry breaking of the fourfold rotation axis with the space group changing from $P4/nmm$ to $Cmme$ (former notation $Cmma$), which is not accompanied by long-range magnetic order \cite{Mc2009a}. The deviation in the inter-atomic Fe--Fe distances along $a$ and $b$ axes due to the orthorhombic distortion is found to be less than 0.5\%, and hence, it is generally believed that the transition is not driven by an instability of the lattice, i.e., a soft phonon mode. Instead, the primary order parameter of the phase transition is considered to be of electronic (nematic) origin \cite{Fer2014} 
related to either electronic spin\cite{Wang2016a}, orbital \cite{Baek2014}, or charge \cite{Mass2016} degrees of freedom. The suggestions include exotic order parameters such as antiferroquadrupolar oder \cite{Yu2015}, stripe quadrupolar order \cite{Zha2017} and collective modes such as a Pomeranchuck instability \cite{Mass2016,Kle2017} of the Fermi surface. Although at ambient pressures, the nematic phase transition is not followed by a long range magnetic order, application of hydrostatic pressure less than 2 GPa seems to induce an elusive, low-moment, spin density wave (SDW) order \cite{Ben2010,Ben2012,Tera2015,Tera2016,Sun2016,Wang2016}. Further, the hydrostatic pressure suppresses the nematic phase \cite{Sun2016,Wang2016,Koth2016} but enhances the superconducting transition temperature $T_c$ from 8.5~K at ambient pressure to 37~K at an optimal pressure of 6 GPa \cite{Miz2008,Med2009,Marg2009,Imai2009} suggesting that the factors enhancing the stability of the nematic phase reduce the superconducting coupling.   
%
%

Nonetheless, even a decade after the discovery of superconductivity in FeSe \cite{Hsu2008}, several aspects of its electronic properties even at ambient pressure, both in the nematic as well as in the superconducting state, still remain unsettled. An early transmission electron microscopy (TEM) study performed by McQueen $et~al.$ indicated that the crystal symmetry below 20 K is likely lower than that of $Cmme$ space group \cite{Mc2009a}. Although the authors did not determine the low-temperature structure, they proposed two possible scenarios which involved an unidirectional distortion in the interatomic Fe--Fe distances either along $a$ or $b$-axis. These symmetry-reducing distortions can lead to a reconstruction of the Fermi surface. In our earlier work, we identified an incipient ordering mode which gave rise to a suppression of the density of states (DOS) in the tunneling spectra obtained by a scanning tunneling microscope (STM), and was associated to the Fe--Fe distortion \cite{Ros2015}. The opening of the partial gap was also identified from the validity of Kohler's scaling of magnetoresistance below 30 K. Alternatively, Kasahara $et~al.$ \cite{Kasa2016} reported an observation of a pseudogap at 20~K based on anomalies found in thermal conductivity and Hall effect measurements and interpreted it as originating from preformed Cooper pairs akin to the case of high-$T_{c}$ cuprate superconductors. In some STM measurements, the suppression of the DOS was not detected at all \cite{Song2011,Spr2017}. Thus, both the presence of the pseudogap as well as its origin require to be clarified. Further, questions such as, how many different superconducting gaps are present \cite{Dong2009,Kha2010,Lin2011,Abd2013,Kas2014,Bour2016,Tek2016,Li2016,Lin2016}, whether the gaps contain accidental nodes \cite{Kas2014,Bour2016,Tek2016,Li2016,Lin2016,Has2018}, and does the order parameter of superconductivity change sign between the hole and electron pockets \cite{Maz2008,Kon2010,Spr2017,Lin2017}, are all controversially discussed in literature.\\ 

In order to check whether some of these controversies can be traced back to the minuscule differences in the samples, we prepared single crystals of FeSe by two different methods, firstly by chemical vapor transport (CVT) using AlCl$_{3}$ \cite{Koz2014,Ros2016}, and secondly by the more popular flux-vapor transport (flux-VT) using a eutectic mixture of KCl and AlCl$_{3}$ \cite{Char2013,Bohm2013}. 
The amount of scattering centers in the samples was quantified by the residual resistivity ratio, RRR.
We show that: (i) In addition to disorder, internal local strain in the crystal can influence the value of $T_{c}$ determined through resistivity measurement. (ii) \sr{From the behavior of Kohler's scaling, two temperature scales are identified, $T^{*} \approx 75$ K and $T^{**} \approx 30$ K. The $T^{**}$--scale is, as also shown by STS measurements, linked to a pseudogap feature.
If the disorder is large, signatures of anisotropic scattering, i.~e. features representing $T^{*}$ and $T^{**}$, can be smeared and even lost}. (iii) The temperature dependence of the electronic part of the specific heat shows a significant difference in samples grown by these two techniques.\\        

\section{Experimental}

\sr{We investigated a total of eight different FeSe samples, five of them grown by CVT (C1--C5) and three by flux-VT method. Among the CVT-grown samples, three were used for transport measurements and the 
remaining two samples (C4,C5) were used exclusively for the STM measurements along with one of the flux-VT samples.}
 

%
\begin{figure}
\centering
	   \includegraphics[clip,width=0.95\columnwidth]{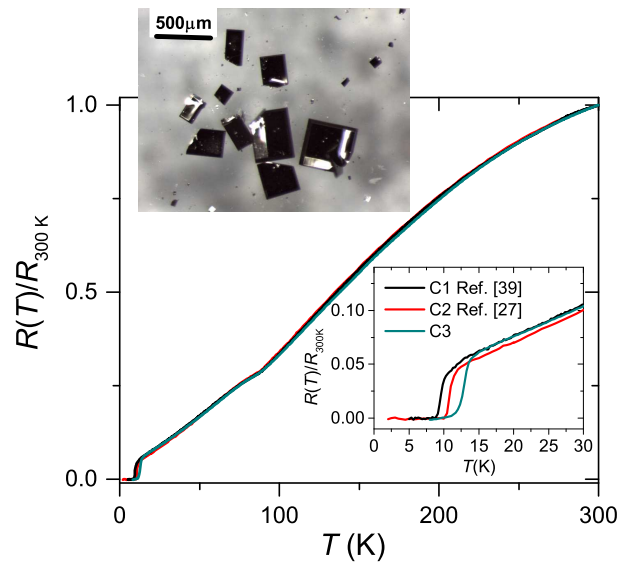}
		\caption{Normalized resistance as a function of temperature measured on three different single crystals grown by CVT. The data for C1 and C2 are from our previous studies published in Refs.~[\onlinecite{Lin2016,Ros2015}], respectively. Samples have similar values for resistivity ratio but display different $T_c$. Left top inset:optical microscopy image of the single crystals grown by CVT. Right bottom inset: $R(T)/R(300~K)$ plot zoomed in to the temperature range 2--30 K.}
	\label{Fig1}
\end{figure}
\begin{figure}
\centering
	   \includegraphics[clip,width=0.95\columnwidth]{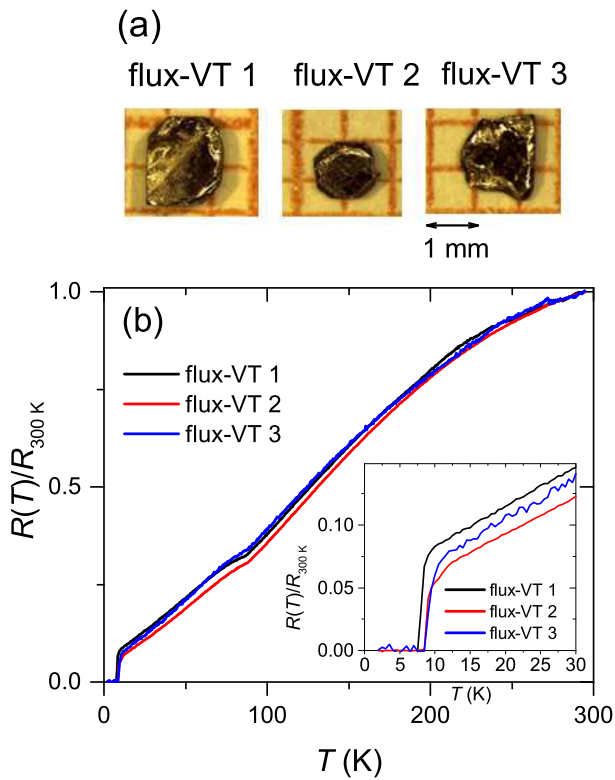}
		\caption{(a) Images of three single crystals grown by flux-VT. (b) Normalized resistance as a function of temperature measured on three different single crystals grown by flux-VT. In the inset, $R(T)/R(300~K)$ plot zoomed in to the temperature range 2--30 K is presented.}
	\label{Fig2}
\end{figure}
\begin{table*}[t!]
\caption{Comparison of the magnetotrasport properties of FeSe single crystals grown by CVT and flux-VT method. Here, RRR is residual resistivity ratio. $T_s$ is the temperature of the structural phase transition. $T^*$ and $T^{**}$ are onset temperatures of anisotropic spin-fluctuations and pseudogap, respectively. $T_{c}^{mid}$ is the superconducting transition temperature obtained from the transition mid point.} \label{tab1}
\begin{ruledtabular}
\begin{tabular}{llllll}
\hspace*{0.28cm} Sample  & RRR & $T_s$ (K)& $T^*$ (K) & $T^{**}$ (K)& $T_{c}^{mid}$ (K)\\
\hline
\hspace*{0.28cm} C1&16.4 &88.5 & not measured & not measured & 9.5 \\
\hspace*{0.28cm} C2& 16.4 &88.5 & 70 & 30 & 10.8  \\
\hspace*{0.28cm} C3& 16.4 &88.5 & not measured & not measured &12.6 \\
\hspace*{0.28cm} flux-VT1&10.2 & 87.2 & not observed & not observed & 8  \\
\hspace*{0.28cm} flux-VT2& 11.4& 87.5 & 70 & 30 & 9  \\
\hspace*{0.28cm} flux-VT3&12.8 & 87.7 & 70 & 30 & 9\\
\end{tabular}
\end{ruledtabular}
\end{table*}
\subsection{Single crystal growth by chemical vapor transport (CVT)}
The single crystals grown by CVT have been used in all our previous studies comprising of transport, specific heat, and STM investigations \cite{Koz2014,Ros2015,Ros2016,Lin2016,Lin2017}. In this method, a quartz ampoule of 10 cm in length and 2 cm in diameter was heated and evacuated before inserting 1 g of 1:1 FeSe powder and 20 mg of AlCl$_{3}$. The latter procedure was conducted inside an argon-filled glove box. The evacuated and sealed ampoules containing the starting materials were inserted horizontally inside a two-zone furnace with a temperature gradient from $T_{2}$= 400$^{\circ}$C to $T_{1}$= 300$^{\circ}$C. A typical growth was carried out for 2-3 months. Finally, the ampoule was quenched in water. The crystals harvested from the cold end of the ampoule were thoroughly washed with ethanol several times to remove traces of AlCl$_{3}$, dried under vacuum and stored in an argon-filled glove box. The typical sizes of the crystals obtained by this method were $400 \times 200 \times 20~\mathrm{\mu m^3}$ and of tetragonal morphology as can be seen in the inset of Fig.~1. In order to obtain larger single crystals (e.g. for specific heat measurements) the growth had to be extended for up to 6--12 months.

\subsection{Single crystal growth by flux vapor transport (flux-VT)}

The use of a eutectic mixture of KCl and AlCl$_{3}$ is a fast and the most popular method of growing FeSe single crystals \cite{Char2013,Bohm2013}. This method is also known to yield high quality single crystals \cite{Kas2014,Bour2016}. In this case, instead of FeSe powder, Fe and Se elemental powders were introduced in the molar ratio 1.05:1 together with a eutectic mixture of the transport reagent AlCl$_{3}$ and flux KCl (molar ratio 2:1) in an evacuated quartz ampoule. The ratio of total amount of Fe and Se powders to (AlCl$_{3}$+ KCl) mixture was kept 1 : 10 in mass. All preparations were performed inside an argon-filled glove box. The subsequent steps were very similar to those of the CVT.
$i.~e.$, the ampoule containing the mixture was evacuated, sealed, and placed horizontally inside a two-zone furnace with a temperature gradient from $T_{2}$= 400$^{\circ}$C to $T_{1}$= 300$^{\circ}$C. The crystal growth was carried out for four weeks. Finally, the ampoule was quenched in water. Crystals were extracted from the cold part of the ampoule. In contrast to the CVT, the single crystals were washed several times using deionized water to dissolve the flux, before they were rinsed in ethanol several times. Finally, crystals were dried under vacuum at room temperature and stored in an argon-filled glove box.

\subsection{Physical properties}
The magnetization $M(T)$ and resistivity $\rho(T)$ measurements were performed using a magnetic as well as physical property measurement systems (MPMS and PPMS, Quantum Design), respectively. The resistivity was measured in the $ab$ plane, for magnetoresistance a magnetic field up to 9 T was applied parallel to the [001] direction of the single crystal. \sr{The electrical contacts were made using gold wires and silver paint. The thicknesses of the three CVT crystals C1, C2, and C3 used in the resistivity measurements were 17, 18 and 20 $\mu$m, respectively and the thicknesses of three Flux-VT crystals flux-VT1, flux-VT2, and flux-VT3 were 0.21, 0.24, and 0.3 mm, respectively}. The specific heat $C_p(T,B)$ was measured down to 0.5~K using a thermal-relaxation method in a PPMS. The scanning tunneling microscopy (STM) and spectroscopy measurements were conducted using ultra-high vacuum (UHV) systems (Omicron Nanotechnology) at base pressures p $\lesssim 2 \times 10^{-9}$~Pa. Two different systems were used to cover the temperature range from 0.35--30 K. In each system, the samples were cleaved $in~situ$ at 20~K \sr{using the so-called post cleave method: A post made of stainless steel is glued onto the FeSe crystal which in turn is mounted on the STM sample plate using a two-component epoxy (H21D, EPO-TEK) before inserting it in to the STM UHV chamber. Inside the UHV chamber, the sample was cooled to 20 K using a flow cryostat and the metal post was knocked off using a manipulator. Subsequently, the sample was inserted into the STM head kept at base temperature of 4.6~K. Since FeSe is a layered compound, cleaving easily exposes a (001) plane with Se-termination.} The tunneling conductance $dI(V)/dV$ was measured directly via lock-in technique. For the measurements, conventional bias settings were used, $i.~e.$, positive bias voltages probe the unoccupied states and negative bias voltages probe occupied states. 
\begin{figure}
\centering
	   \includegraphics[clip,width=0.9\columnwidth]{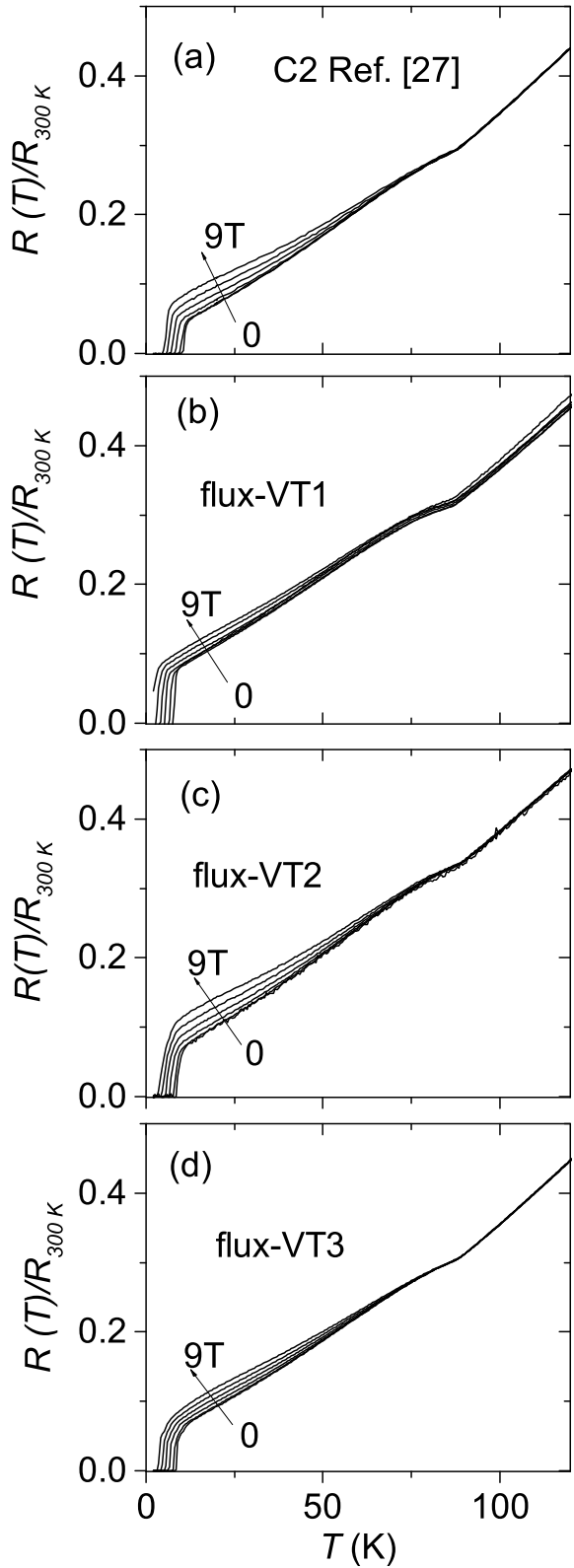}
		\caption{The temperature dependence of normalized resistance $R(T)/R(300~K)$ measured in different magnetic fields for (a) C2 taken fron Ref.[\onlinecite{Ros2015}], and (b)-(d) crystals grown by flux-VT. The magnetic field was applied parallel to the $c$-axis.}
	\label{Fig3}
\end{figure}
\section{Results and discussion}

\subsection{Residual resistance ratio RRR}

The amount of scattering centers in a metallic system can be estimated semi-quantitatively by inspecting the value of the residual resistance ratio (RRR) by taking the ratio of the resistances at room temperature and at $T\rightarrow 0$.  Since FeSe is superconducting below 8.5 K, we took the resistance ratios at temperatures 300 K and 15 K,  i.e., RRR=$R_{300\mathrm{K}}/R_{15\mathrm{K}}$, to avoid any uncertainty in the resistance value at $T\rightarrow 0$ due to an extrapolation. However, in the case of FeSe, below $T_s$, the value of RRR may be influenced not only by the scattering centers induced by impurities but also from unavoidable twin boundaries present in the orthorhombic phase of the crystal \cite{Tek2016,Lin2016,Kno2015}. 
In Fig. 1, $R(T)/R(300~K)$ plots of three different FeSe crystals grown by CVT are presented. Remarkably, down to 15 K, all the three crystals display a similar behavior with RRR $\approx16.4$, indicating that they contain nearly a similar amount of scattering centers. Nonetheless, the crystals displayed significantly different superconducting transition temperatures with the value of $T_c^{mid}$ (obtained from the transition mid point) varying from 12.6 K to 9.5 K. However, as our previous reports \cite{Koz2014,Lin2016} show, the bulk $T_c$ obtained from the heat capacity and magnetization measurements on these crystals were found to be 8.5 K. One possible scenario for this discrepancy could be internal strain locally present in the crystal, which may induce superconducting percolation paths. Thus, the $T_c$ obtained from the resistivity measurement in FeSe can be influenced both by concentration of scattering centers as well as strain in the crystals.    

In Fig. 2(a), images of three single crystals grown by flux-VT method are presented. Fig. 2(b) shows $R(T)/R(300~K)$ plots of these three crystals. The RRR values of these crystals were found to 10.2 for flux-VT1, 11.4 for flux-VT2, and 12.8 for flux-VT3. These values are significantly lower than those measured in the crystals grown by CVT. The flux-VT method is known\cite{Kas2014,Bour2016,Kno2015,Dong2009,Chen2017} to produce single crystals with values of RRR varying from 23 to 5. As pointed out by Kn\"oner $et~al.$ \cite{Kno2015}, in addition to the initial inherent disorder incorporated in the crystal during the growth process, cooling the crystals through the structural transition temperature $T_s$ introduces different twin states in the crystals. Moreover, the in-plane anisotropy may also contribute to different behavior of resistivity measured in the $ab$ plane.     

As mentioned above, the $T_c$ of FeSe crystals can be influenced by both internal strain and disorder. \sr{Hence, the $T_c$ values obtained from the resistivity measurements cannot be directly correlated with the RRR.} Nonetheless, \sr{we would like to point out that among the crystals studied here, the crystal with the lowest value of RRR showed the lowest $T_c$ (see Table I) as well  as the lowest value of the upper critical field $H_{c2}$ (Fig. 3(b)).} In Fig. 3, $R(T)/R(300~K)$ of crystals grown by both CVT and flux-VT methods measured in magnetic fields up to 9~T are presented. The C2 crystal grown by CVT (Ref. [\onlinecite{Ros2015}]) showed a $T_c^{mid} \approx$ 10.8~K as can be seen in Fig. 3(a).
Whereas flux-VT2 and flux-VT3 samples displayed a $T_c^{mid} \approx$ 9~K, flux-VT1 sample with the lowest value of RRR among the samples studied here showed $T_c^{mid} \approx$ 8~K. Further, in an applied magnetic field of 9~T, the superconducting transition of sample flux-VT1 (Fig. 3b) is incomplete at 2 K, suggesting that this crystal has also a lower $H_{c2}$. \sr{In Table I, the values of RRR, $T_s$, and $T_c^{mid}$ obtained from the resistivity measurements are compiled for comparison.}

\subsection{Kohler's scaling} 
In an intention to check whether there is a possible sample dependency in the signature of the incipient ordering mode previously reported by us in FeSe grown by CVT \cite{Ros2015}, we measured the magnetoresistance (MR) of all the three flux-VT grown crystals. From the scaling behavior of MR known as Kohler's rule, it is possible to identify if there is any anisotropic quasiparticle scattering at some points of the Fermi surface. According to this rule, if the scattering rates for charge carriers are equal at all points of the Fermi surface, $i.~e.$, if the scattering rates are isotropic, the MR =  $[\rho(H)-\rho(0)]/\rho(0)$ should scale with magnetic field $H$ as an arbitrary function $\mathcal{F}[H/\rho(0)]$ regardless of the topology of the Fermi surface. In FeSe, with imperfect nesting of electron and hole Fermi surfaces, hot spots and cold parts with short and long life times, respectively, are expected \cite{Bre2014}. The MR measurements \cite{Ros2015,Su2016,Ter2016} of single crystalline FeSe have shown that the Kohler's rule is violated at temperatures $T < T_{s}$, indicating the presence of anisotropic scattering on the Fermi surface. Surprisingly, however, the Kohler's rule becomes valid below $T^{**} \approx$ 30 K \cite{Ros2015,Su2016,Ter2016}, which suggests an opening up of a partial gap \cite{Ros2015,Su2016} possibly related to a pseudogap formation \cite{Kasa2016}.
In Fig. 4 (a-c), we show the Kohler's scaling for the flux-VT crystals. The signature of anisotropic scattering rates and the subsequent opening of the gap at $T^{**}$ is absent in flux-VT1 sample with the lowest value of RRR. The crystals flux-VT2 and flux-VT3, although only slightly better in quality than flux-VT1, display the signs of anisotropic scattering below $T_{s}$, and a subsequent recovery of Kohler's scaling (Figs. 4 (b) and (c)) below $T^{**}$. \sr{The results of the scaling behavior of different samples are summarized in Table I.} Thus, if the disorder in the crystals is high, telltale signs of temperature dependent modifications of the Fermi surface become smeared and the scattering rate appears isotropic.
\begin{figure}[t]
\centering
	   \includegraphics[clip,width=0.9\columnwidth]{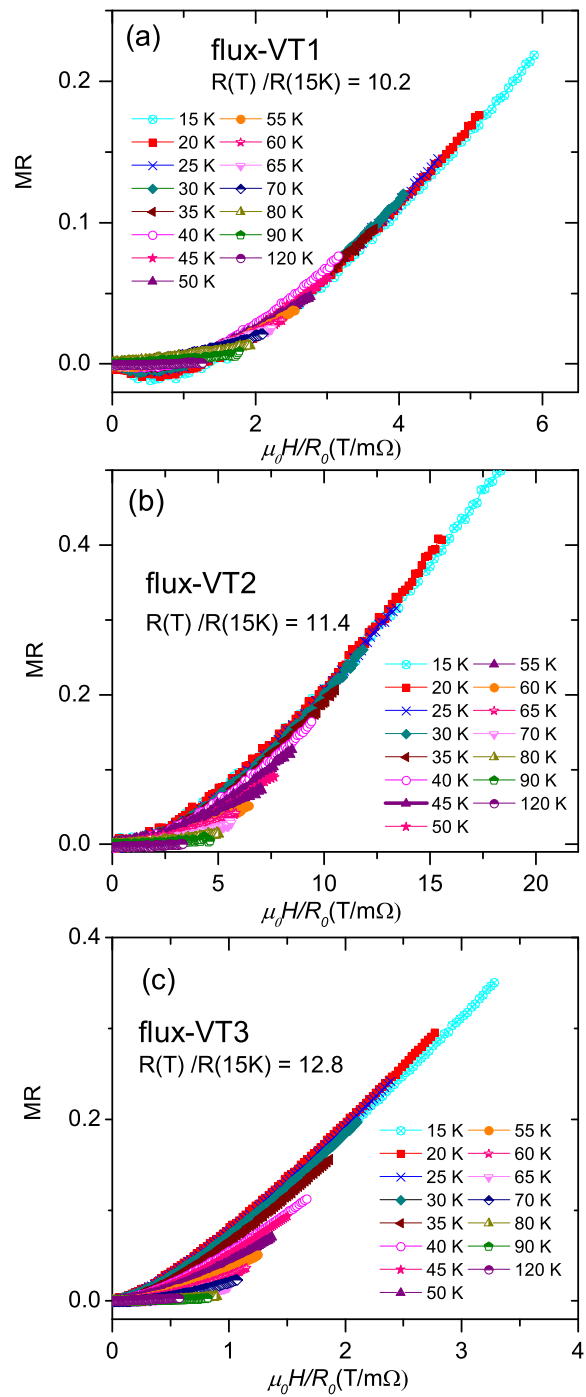}
		\caption{Kohler plots in the temperature range 15 -120 K for the crystals grown by the (AlCl$_{3}$+KCl) method. The scaling behavior is strongly dependent on the sample quality.}
	\label{Fig4}
\end{figure}
\begin{figure}[t]
\centering
	   \includegraphics[clip,width=0.9\columnwidth]{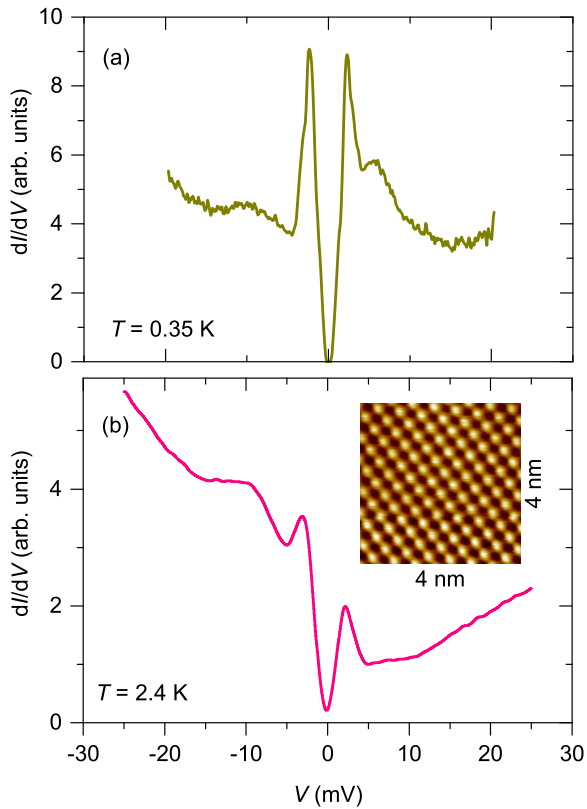}
		\caption{(Color online) Scanning tunneling spectroscopy measurements performed on crystal C4 grown by CVT. (a) at 0.35 K with modulation voltage $V_{mod}$=0.05 mV, also reported in Ref.[\onlinecite{Lin2016}] and (b) at 2.4 K, $V_{mod}$=0.3 mV. Inset displays a topography on an area of 4 nm $\times$ 4 nm. The tunneling parameters used for the topography measurements were as follows. The current set point $I_{sp}$=100 pA, the bias voltage $V_{b}$=10 mV.}
	\label{Fig5}
\end{figure}
\begin{figure}[t]
\centering
	   \includegraphics[clip,width=0.9\columnwidth]{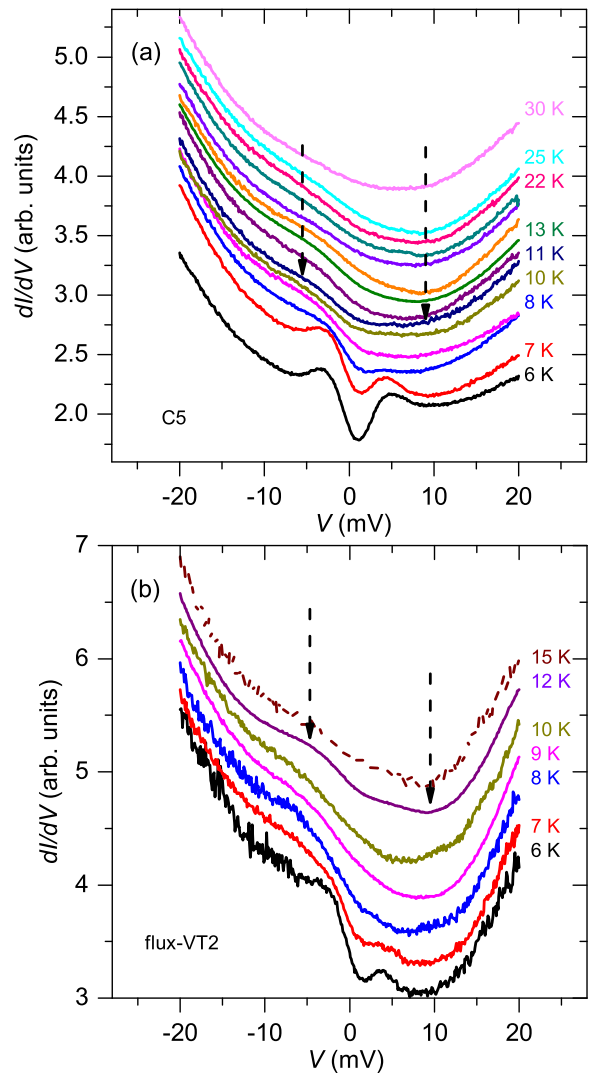}
		\caption{(Color online) Scanning tunneling spectroscopy measurements performed at different temperatures on FeSe single crystals (a) sample C5, with tunneling parameters current set point $I_{sp}$= 800 pA, and bias voltage $V_{b}$= 20 mV  (b) flux-VT2 sample with $I_{sp}$= 500 pA, and bias voltage $V_{b}$= 20 mV. The modulation voltage $V_{mod}$= 0.1 mV was used for both measurements.}
	\label{Fig6}
\end{figure}

\begin{figure}[t]
\centering
	   \includegraphics[clip,width=0.9\columnwidth]{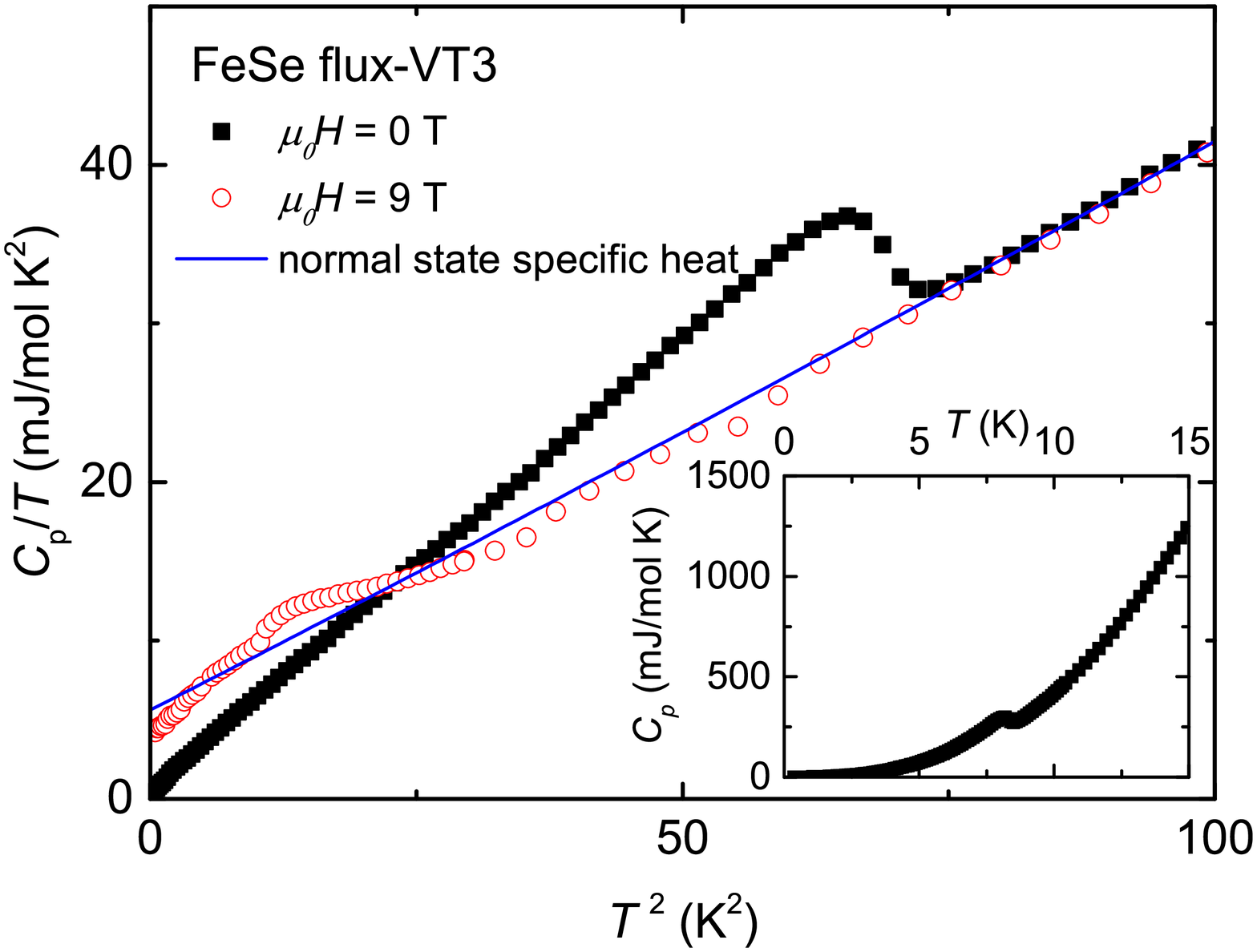}
		\caption{(Color online) Specific heat divided by temperature, $C_p/T$ vs $T^{2}$, measured in zero and a magnetic field of 9 T for FeSe flux-VT3. The solid line represent the normal state specific heat $C_n(T)$. The inset shows $C_p(T)$ of the same sample.}
	\label{Fig7}
\end{figure}

\subsection{Scanning tunneling spectroscopy}

In Fig. 5, scanning tunneling spectroscopy conducted on an FeSe single crystal (Sample C4) grown by the CVT is presented. All the spectroscopic measurements were performed on atomically resolved Se-terminated surfaces. A typical topography of area 4 nm $\times$ 4 nm is shown in the inset of Fig. 5 (b). The tunneling spectrum measured at 0.35~K displays a superconducting gap with coherence peaks appearing at $\approx$2.35~mV. In Ref. \onlinecite{Lin2016}, a detailed estimation of the superconducting gap structure can be found. One of the striking feature of these spectra is that they display an asymmetric background tunneling conductance with respect to zero bias voltage. This strongly suggests an uncompensated nature of the occupied and unoccupied states. With increasing temperature (Figs.5(b) and 6), the asymmetry appears even in the coherence peaks. This behavior indicates that a complete superconducting coherence in all involved bands is achieved only below about 2~K, which is consistent with the conclusions drawn from the specific heat measurement \cite{Lin2016} as discussed below in subsection D.

Now we address spectroscopic features observed at even higher temperatures, particularly around $T^{**}$ discussed already in section \textrm{III}B. In our previous publication \cite{Ros2015}, we reported a signature of an incipient ordering observed at $T^{**}$ in scanning tunneling spectroscopy as a weak suppression of the local density of states (LDOS) in the tunneling spectra. However, such a depression in LDOS was not observed in other STM measurements on FeSe \cite{Song2011,Spr2017}. Therefore, we decided to measure the temperature dependence of tunneling conductance $dI(V)/dV$ on two additional FeSe crystals, one grown by CVT (sample C5) and the other by the flux-VT (sample flux-VT2) method. These data are presented in Fig.~6(a) and (b), respectively.
In the spectra measured at 6~K, the superconducting gap is visible including the corresponding coherence peaks. 
%
Besides the asymmetry, a second, more subtle, feature found in $dI(V)/dV$ curves is the continued suppression of the LDOS (marked by arrows in Fig. 6) even at temperatures above $T_c\approx$ 8.5 K. The feature is more prominent in CVT grown sample. A shallow minimum close to the Fermi level can be tracked up to 22~K in the CVT grown FeSe and up to 15~K in Flux-VT2 crystal. 
Thus, these new measurements are consistent with our previous report\cite{Ros2015}. \sr{Further, our tunneling spectra presented in Figs. 6 (a) and (b) indicate that the suppression of LDOS at $ T > T_c$ is pertinent both to the CVT and flux-VT crystals.}
In the high-$T_c$ cuprates, which are single band unconventional superconductors, such a suppression of the LDOS is attributed to a pseudogap precursor of the superconducting gap \cite{Ren1998}. In addition to our earlier report \cite{Ros2015} of a suppression of the LDOS even well above $T_{c}$ in FeSe, giant superconducting fluctuations and pseudogap behavior below $T$ = 20~K has been reported \cite{Kasa2016} based on thermal conductivity and Hall effect measurements. Further, as described above in the section of Kohler's scaling and in Refs.~\onlinecite{Ros2015,Su2016,Ter2016}, partial opening of the gap has been identified also by observing the validity of scaling behavior below  $T^{**} \approx$ 30 K. 


\subsection{Specific heat}
In the following, we discuss the signature of multigap superconducting behavior found in the temperature dependence of specific heat $C_p(T)$ measurements of FeSe. In addition to the jump of $C_p(T)$ at $T_c \approx$ 8.5~K, a small hump was observed at $T\approx 1.5$ K \cite{Lin2016,Sun2017a,Sun2017b}. This type of feature typically appears in multigap superconductors such as MgB$_{2}$ and Lu$_{2}$Fe$_{3}$Si$_{5}$ \cite{Bouquet2001,Naka2008}. Alternatively, in FeSe, this hump was also interpreted as originating from a spin-density wave order below 1.5 K \cite{Chen2017}. In general, $C_p(T)$ at $T\rightarrow$ 0 is one of the powerful methods to investigate whether the superconducting gap structure contains nodes. In the case of a fully gapped ($s$--wave) superconductors, $C_p(T)$ follows an activated temperature dependence, $\exp (-\Delta/T)$ at $T\ll T_c$ \cite{Mat2006}. In nodal superconductors, on the other hand, low energy excitations remain finite at $T\rightarrow$ 0 and $C_p(T)$ is expected to follow a linear or quadratic behavior depending on the topology of nodes \cite{Mat2006}. In the case of FeSe, such an analysis of $C_p(T)$ at $T\rightarrow$ 0 may be hampered owing to a multigap behavior. Moreover, if the feature related to the small gap is smeared due to disorder, the low-temperature $C_p(T)$ might mimic a nodal behavior as we show below.

\begin{figure}[t]
\centering
	   \includegraphics[clip,width=0.9\columnwidth]{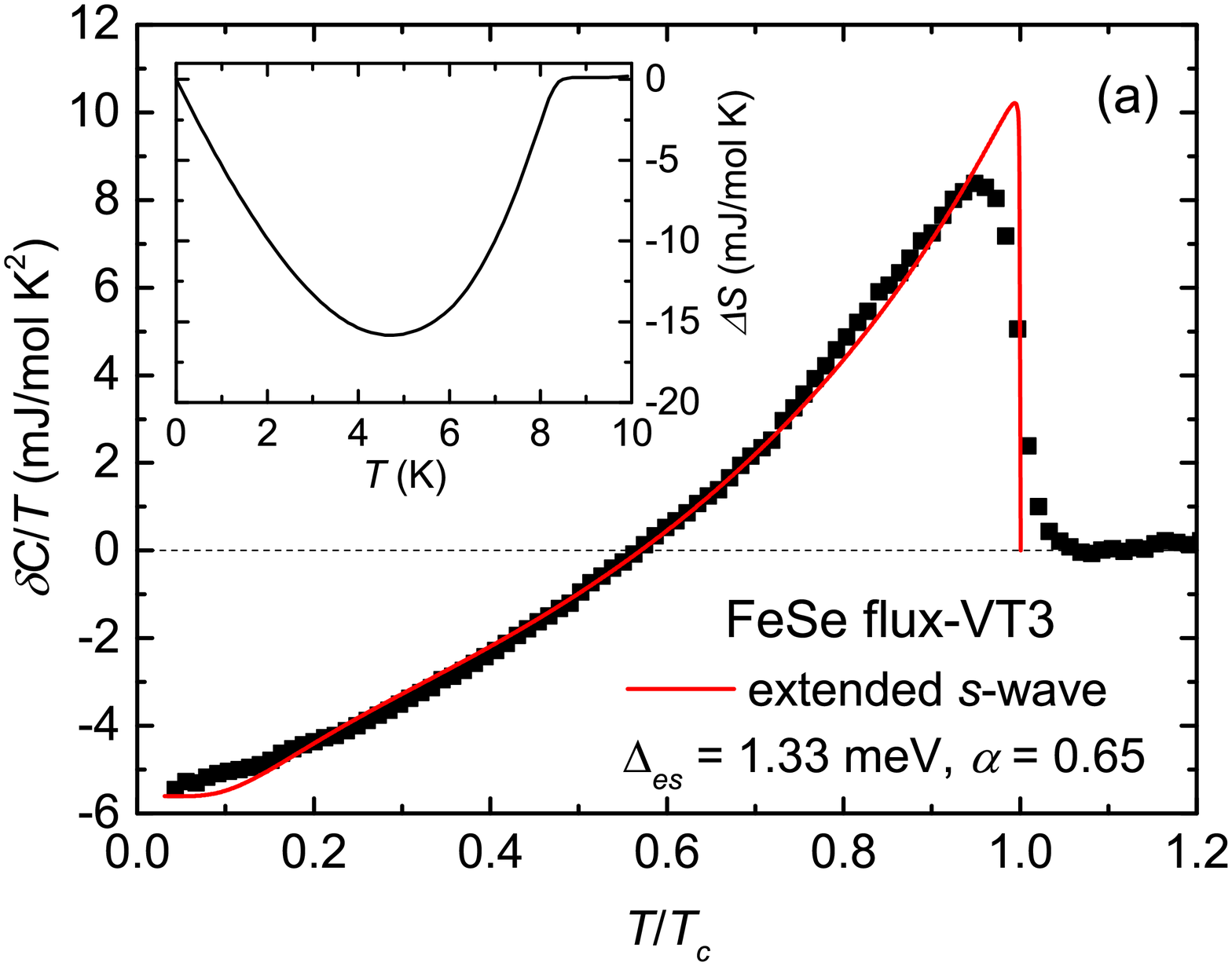}
		 \includegraphics[clip,width=0.9\columnwidth]{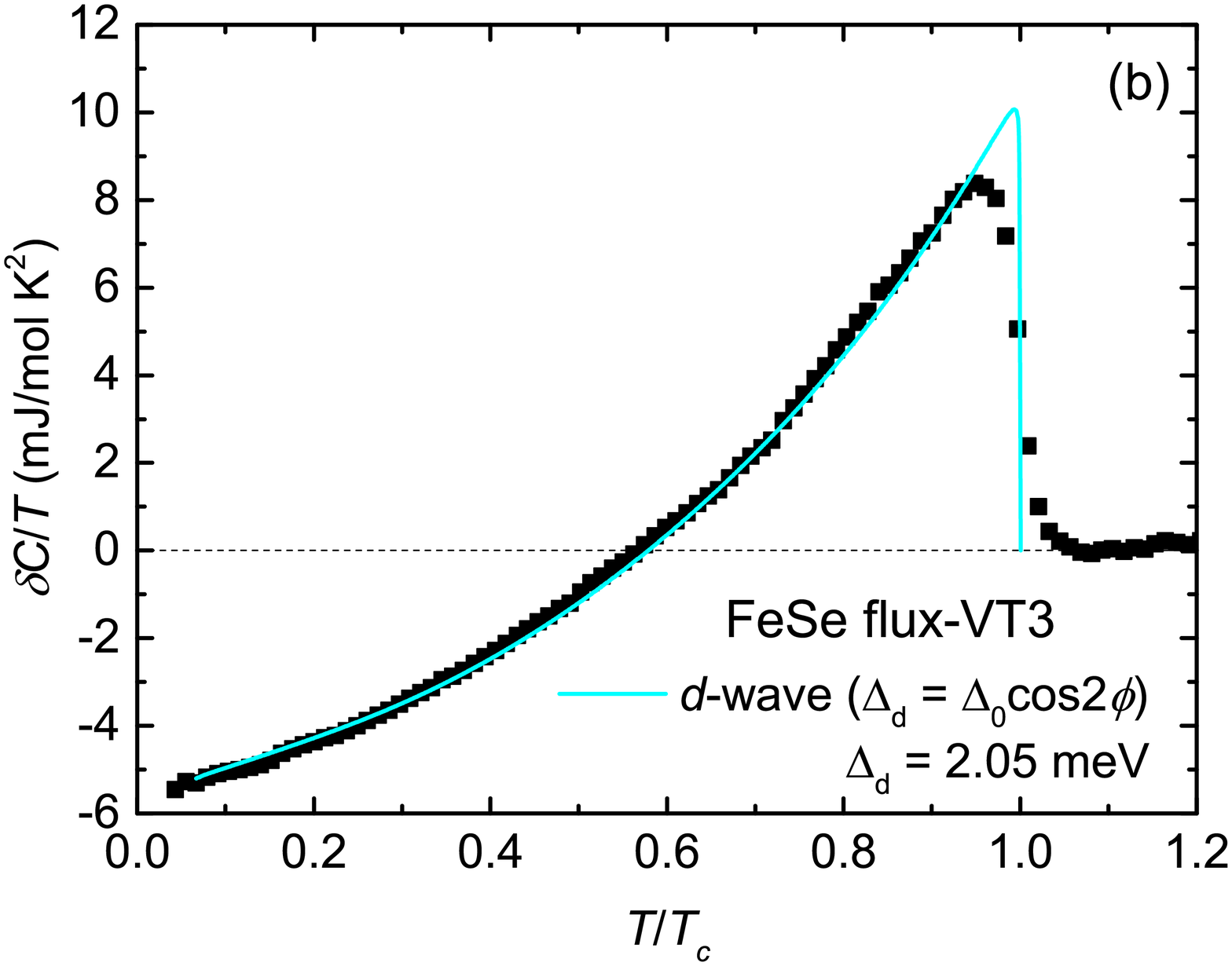}
		\caption{(Color online) The excess electronic specific heat contribution in the superconducting state $\delta C(T)/T$ as a function of reduced temperature $T/T_c$ for flux-VT3 sample with fittings (a) an extended $s$-wave and (b) a $d$-wave model. The inset shows the entropy 
conservation required for a second-order phase transition.
}
	\label{Fig5a}
\end{figure}
 
The specific heat $C_p(T,B)$ of FeSe sample flux-VT3 was measured with the magnetic field $B$ applied parallel to the [001] direction of the single crystal. The zero-field $C_p/T$ vs $T^{2}$ plot between 0.35 and 15~K is presented in Fig. 7. The data displays a $\lambda$-like transition at $T_{c} = 8.32(1)$~K. In Fig. 8, the excess electronic specific heat contribution in the superconducting state, $\delta C(T)/T$, is plotted as a function of reduced temperature. $\delta C(T)$ was calculated by subtracting the specific heat contribution in the normal state, $C_n(T)$, from the total specific heat: $\delta C(T) = C_{p}(T, B = 0) - C_{n}(T)$. The $C_{n}$ below 10~K was obtained by $C_{n}(T) = \gamma_{n}T + C_{lat}(T)$, where $\gamma_{n}T$ is the normal electronic contribution and $C_{lat}(T) = \beta_{3} T^{3} + \beta_{5} T^{5}$ represents the phonon contribution. A fit to $C_p(T, 0\mathrm{T})/T$ in the temperature range 9--13~K yields $\gamma_{n} = 5.61$~mJ/mol K$^2$, $\beta_{3} = 0.342$~mJ/mol~K$^4$, and $\beta_{5} =$1.67$\times 10^{-4}$~mJ/mol~K$^6$. The Debye temperature $\theta_{\texttt{D}}$ calculated from $\beta_{3}$ is 225~K. The insets in Figs. 8(a) and (b) illustrate the satisfaction of entropy conservation $\Delta S = \int_{0}^{T_{c}}(\delta C/T)dT$ justifying the validity of the parameters used to fit $C_{n}(T)$. 
The normalized specific-heat jump at $T_{c}$, $\Delta C/\gamma_{n}T_{c}$, is estimated to be 1.80, which is slightly larger than the weak-coupling value 1.43 of Bardeen-Cooper-Schrieffer (BCS) theory \cite{Bar1957}.
\begin{figure}[t]
\centering
	   \includegraphics[clip,width=0.9\columnwidth]{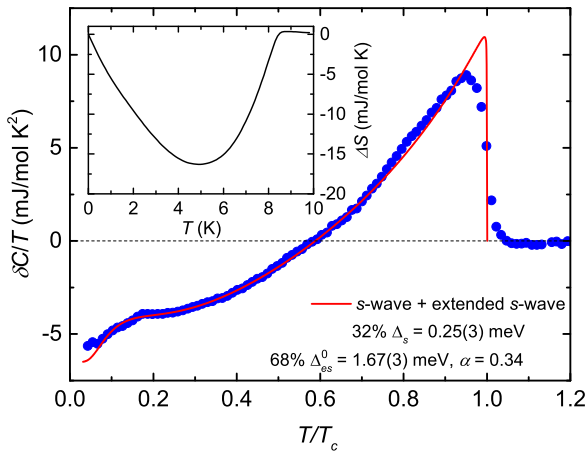}
		 \caption{(Color online) The excess electronic specific heat contribution in the superconducting state $\delta C(T)/T$ as a function of reduced temperature $T/T_c$ for a CVT grown FeSe sample. 
		For the fitting, a two-band model with an $s$--wave and an extended  $s$--wave gap was used\cite{Lin2016}.
		The inset shows the entropy 
conservation required for a second-order phase transition.}
		\label{Fig5a}
\end{figure}

For the sake of scrutinizing the superconducting order parameter, the excess electronic specific heat contribution in the superconducting state $\delta C(T)/T$ was fitted to the one-band BCS equation \cite{Bar1957}. The fitting procedure followed here was very similar to that described in our previous publication on superconducting gap structure obtained for FeSe grown by CVT reported in Ref. \onlinecite{Lin2016}. The data were fitted with two different models, an extended $s$-wave gap $\Delta(T, \theta) = \Delta_{es}^0(T)(1 + \alpha~\mathrm{cos} 4\theta)$ (Fig. 8(a)) and a $d$-wave gap $\Delta_{d}(T, \theta) = \Delta_{d}^0(T)\mathrm{cos} 2\theta$ (Fig. 8(b)). Here, $\alpha$ and $\theta$ represent the gap anisotropy and polar angle, respectively. For the extended $s$-wave model, the best fit was found for the values $\Delta_{es}^0$=1.33 meV and $\alpha$=0.65. For the $d$-wave model, a best fit was obtained for $\Delta_{d}^0$=2.05 meV. As can be seen in Fig (b), the single gap $d$-wave model is sufficient to represent the $\delta C(T)/T$ data better at low temperatures. 

In Fig. 9, $\delta C(T)/T$ data of FeSe grown by CVT published in Ref. \onlinecite{Lin2016} is presented for comparison. In addition to a jump at $T_{c}$,
a broad shoulder below 2~K was observed in $\delta C(T)/T$. This feature is also reported for FeSe crystals in Refs. \onlinecite{Chen2017,Sun2017a,Sun2017b}, which is not present in the $\delta C(T)/T$ data of flux-VT3 sample. The solid line in Fig. 9 was obtained by fitting a two band BCS model, a detailed description of the fitting procedure can be found in Ref.~\onlinecite{Lin2016}. In addition to an extended $s$--wave model, a small isotropic gap was found to be necessary to describe the shoulder in $\delta C(T)/T$ found below 2~K. As mentioned above, the FeSe single crystals grown by CVT typically showed an RRR of 16.4, which is significantly higher than that of our flux-VT3 sample. The signature of the smaller superconducting gap \cite{Lin2016,Sun2017a,Sun2017b} in the $\delta C(T)/T$--data of flux-VT3 sample is likely smeared by the scattering produced by impurities. Hence, the low-temperature electronic specific heat of flux-VT3 sample emulates the behavior expected for nodal superconductors.   

\subsection{Discussion}

FeSe displays a reduction of the rotation symmetry from fourfold ($C_4$) to twofold ($C_2$) at
$T_{s} \approx$ 87~K as a result of a change of crystal structure from a tetragonal to orthorhombic phase \cite{Mc2009a}. The order parameter driving this transition is believed to be of an electronic origin \cite{Fer2014}. Upon further decreasing the temperature, anomalies were identified, at temperatures $T^* \approx$ 75--70~K and $T^{**}\approx$ 30--20~K. The temperature $T^*$ was detected in magnetic susceptibility \cite{Ros2015}, magnetoresistance \cite{Ros2015}, Hall effect \cite{Ros2015,Kas2014}, and $1/T_{1}T$ relaxation time measured in nuclear magnetic resonance (NMR) experiments \cite{Imai2009}. Since all these techniques require magnetic field for measurements, the measured properties very likely manifest a response of the underlying electronic spin. Therefore, $T^*$ can be considered as an onset temperature of enhanced anisotropic spin-fluctuations, which induce momentum-dependent anisotropy in the scattering rates over the Fermi surface. This is also reminiscent of the behavior found in non-Fermi liquids.  However, the physical mechanism occurring at $T^{**}$ is much less straight-forward. At this temperature, scattering rates over the Fermi surface become once again isotropic, as depicted by the validity of Kohler's rule\cite{Ros2015,Su2016,Ter2016} below $T^{**}$, see Fig. 4. The Hall coefficient, thermal conductivity, and scanning tunneling spectroscopy measurements indicate a suppression of electronic DOS \cite{Ros2015,Kasa2016}. The symmetry is likely lower than that described by the $Cmme$ space group \cite{Mc2009a}. At these low temperatures, a one-dimensional bond-order-wave has also been suggested by an ARPES experiment \cite{Wat2016}. All these anomalies in the physical properties indicate a precursor state occurring at $T^{**}$, which sets the stage for a strongly anisotropic superconducting gap below $\approx$ 8.5~K. Whether this precursor state competes with, or promotes, superconductivity is an open question to be investigated.
        
As a consequence of the underlying $C_2$ symmetry of the electronic system, the superconductivity in FeSe is quasi one-dimensional, $i.~e.$, the superconducting gap structure displays a strong anisotropy, which was first identified through an analysis of excess electronic specific heat \cite{Lin2011}. A strong anisotropy in the superconducting gaps has also been confirmed in a momentum-resolved  Bogoliubov quasiparticle interference (BQPI) experiment \cite{Spr2017}. These experiments also showed two anisotropic superconducting gaps, nearly of equal magnitude residing on a hole band at the $\Gamma$ point and an electron band at the $M$ point.  An orbital-selective Cooper pairing in FeSe has been proposed to explain the gap anisotropy \cite{Spr2017}. Further, experiments such as phase-resolved BQPI imaging \cite{Spr2017} and observation of a non-magnetic impurity induced bound-states in the tunneling spectra \cite{Lin2017} indicated that the order parameter is changing sign between the hole and electron bands. However, bulk measurements such as specific heat \cite{Lin2011,Lin2016,Sun2017a,Sun2017b}, thermal conductivity\cite{Bour2016}, and penetration depth \cite{Tek2016}experiments found indications of a smaller gap of magnitude 0.2--0.6 meV.  Since, these bulk measurements are not momentum-resolved, the location of the smaller gap is yet to be identified. Recent specific heat measurement even suggested three superconducting gaps \cite{Sun2017b}. It is likely that some of the bands are elusive to certain types of spectroscopic measurements \cite{Spr2017}.\\     

\section{Conclusions}
We have shown by using resistivity, magnetoresistance, scanning tunneling spectroscopy, and specific heat measurements on multiple samples, also grown by different techniques, that some of the physical properties controversially discussed in literature might be related to the quality of the samples. That is, if the residual resistivity ratio of the sample is lower, which in our study holds for the flux-grown specimens, subtle features representing a suppression of the density of states commencing at $T^{**}\approx$ 20--30~K and the smaller superconducting gap may be lost due to scatterings induced by disorder. Identifying those features which are strongly sample-quality dependent will therefore clearly contribute to a clarification of the mechanism of superconductivity in FeSe. 

\sr{During the review stage of this manuscript, we found a paper published on the arXiv, which also identifies the temperature scales $T^{*}$ and $T^{**}$ using muon spin rotation ($\mu$SR) experiments \cite{Gre2018}.} 

\section{Acknowledgments}
We thank S.-L. Drechsler, and U. K. R\"o{\ss}ler for fascinating discussions.
Financial support from the Deutsche Forschungsgemeinschaft (DFG) within the Schwerpunktprogramm SPP1458 is gratefully acknowledged. L.J. thanks the Alexander-von-Humboldt foundation for financial support.


\end{document}